\documentclass[journal]{IEEEtran}
\usepackage{graphicx}
\usepackage{subfigure}
\usepackage{amsfonts}
\usepackage{amsmath}

\usepackage{tikz}
\usepackage{xcolor}
\usepackage{eso-pic}

\usepackage{bm}  % for bold

\usepackage{booktabs} % to regulate the space between the rows in table
\usepackage{multirow}  % for tables in this article
\usepackage{threeparttable}

\usepackage[colorlinks, 
			linkcolor=black,  % table gram ref
			anchorcolor=black,  % cite
			citecolor=black,  % cite
			urlcolor=blue]{hyperref}  % url
\usepackage{verbatim}
\usepackage{algorithmic}
\usepackage{algorithm}
\usepackage{lipsum}

\newcommand{\RNum}[1]{\uppercase\expandafter{\romannumeral #1\relax}}

\begin{document}
%\watermark{45}{10}{Twistfatezzz}
% at, but, by, for, in, nor, of, on, or, the, to and up
\title{Multi-Lead ECG Classification via an Information-\ Based Attention Convolutional Neural Network}
%\title{Towards End-to-End ECG Classification with Raw Signal Extraction and Deep Neural Networks}

\author{Hao~Tung,~\IEEEmembership{}
       Chao~Zheng,~\IEEEmembership{}
       Xinsheng Mao,
        and~Dahong~Qian,~\IEEEmembership{}% <-this % stops a space
\thanks{H. Tung is with the Institutes of Biomedical Engineering, Shanghai Jiaotong university, China, (e-mail: twistfatezzz@sjtu.edu.cn).}% <-this % stops a space
\thanks{Chao Zheng and Xinsheng Mao are with Shukun (Beijing) Network Technology Co.,Ltd.}
\thanks{D.H. Qian is with the Institutes of Medical Robotics, Shanghai Jiaotong University, China, (e-mail: dahong.qian@sjtu.edu.cn)}% <-this % stops a space
\thanks{Key research and development project in Zhejiang Province, China: ``Wearing human body physiological parameter monitoring technology and equipment research and development -- 5-parameter wearable device and cloud platform based on body area network'', Grant No. 2017C03029.}
%\thanks{Manuscript received; revised. }
}

\markboth{}%
{xx \MakeLowercase{\textit{et al.}}: Multi-lead ECG Classification via a Novel Informatics-based Attention Convolutional Neural Network}
\maketitle

% 302 words
\begin{abstract}Objective: A novel structure based on channel-wise attention mechanism is presented in this paper. Embedding with the proposed structure, an efficient classification model that accepts multi-lead electrocardiogram (ECG) as input is constructed. Methods: One-dimensional convolutional neural networks (CNN) have proven to be effective in pervasive classification tasks, enabling the automatic extraction of features while classifying targets. We implement the Residual connection and design a structure which can learn the weights from the information contained in different channels in the input feature map during the training process. \iffalse To demonstrate the beneficial properties of our proposed structure in the ECG classification task, \fi An indicator named mean square deviation is introduced to monitor the performance of a particular model segment in the classification task on the two out of the five ECG classes. The data in the MIT-BIH arrhythmia database is used and a series of \iffalse subject-oriented \fi control experiments is conducted. Results: Utilizing both leads of the ECG signals as input to the neural network classifier can achieve better classification results than those from using single channel inputs in different application scenarios. Models embedded with the channel-wise attention structure always achieve better scores on sensitivity and precision than the plain Resnet models. The proposed model exceeds the performance of most of the state-of-the-art models in ventricular ectopic beats (VEB) classification, and achieves competitive scores for supraventricular ectopic beats (SVEB). Conclusion: Adopting more lead ECG signals as input can increase the dimensions of the input feature maps, helping to improve both the performance and generalization of the network model. Significance: Due to its end-to-end characteristics, and the extensible intrinsic for multi-lead heart diseases diagnosing, the proposed model can be used for the real-time ECG tracking of ECG waveforms for Holter or wearable devices. 
% Codes is available at \url{https://github.com/twistfatezz} 
\end{abstract}


%\vspace{-5pt}  % adjust failure

% \fi
\IEEEpeerreviewmaketitle

\section{Introduction}
% ----- ECG origin----- %
\IEEEPARstart{I}{n} 1903, Willem Einthoven recorded the first clear human ECG, opening a new era of clinical applications of ECG. Over the past 100 years, thousands of ECG workers, cardiologists and biomedical engineers have laid the foundation, exploration and innovation to make world-famous contributions to the protection of human health and the saving of countless lives. 
% ----- ECG is practical and powerful ----- %
An electrocardiogram is a simple, inexpensive, easy-to-follow observation and an important clinical support program that contains a wealth of information describing the state of cardiac electrophysiology. Electrocardiography is of great value in the diagnosis of abnormal origins of cardiac activation and conduction abnormalities. Arrhythmia describes a group of important diseases among cardiovascular diseases. It can occur alone or in conjunction with other cardiovascular diseases. 
\iffalse Atrial or ventricular arrhythmias can also induce or media other kind of diseases such as cardiomyopathy, which may lead to rapid heart failure \cite{24_3}.\fi 
Many rapid and slow arrhythmias can be diagnosed by ECG, such as various premature beats, supraventricular atrial fibrillation, atrial flutter, atrial fibrillation and so on.

% -------- xxxx ------------ %
However, only experienced cardiologists can easily distinguish between abnormal heartbeats and normal sinus rhythms by observing an electrocardiogram, and it is unlikely that patients with arrhythmia will be monitored by a professional doctor in real time. Standard 12-lead ECGs that are very effective for the diagnosis of multiple heart diseases are often not a means of long-term detection. Although an improved lead configuration facilitates the patient's basic activities (changing the limb lead to the chest side), the signal changes significantly due to changes in the position of the measuring electrode, and some related studies have shown that this placement changes the electrocardiogram's clinical specificity \cite{z_6}. Furthermore, the ECG features of certain arrhythmias are subject-specific and closely related to the physical state and the physiological factors of the subject \cite{2} \cite{21_1}. Differences in ECG measuring instruments and various measurement environments both pose challenges to the task of automatically detecting ECG using computers. 

% ----- shortcomings of "2-stage" ----- %
Early researchers used the ``two-stage'' idea of feature extraction and classification to do much research work on the diagnosis of arrhythmia: using structural and statistical knowledge of the ECG signal classified by a hidden Markov model \cite{26_6}, constructing filter banks that decomposes the ECG into sub-bands with uniform frequency bandwidths \cite{26_3}, classifying the ECG with its morphological information and timing information by a feed-forward multilayer perceptron neural network \cite{26_2}, developing a statistical model towards ECG morphology with intervals features \cite{19_1}, extracting ECG's higher order statistics and Hermite coefficients classified by a support vector machine (SVM)\cite{21_1}, developing morphological and temporal features classified by a multilayer perceptron \cite{z_2}, applying multi-resolution wavelet transform features with SVM classifier \cite{z_3}. However, as mentioned in \cite{19_4}\iffalse and \cite{12_1}\fi, due to the complex and variable application scenarios of the algorithm, the generalization performance of the two-stage algorithms remains to be proven. 

% ----- 1-lead works ----- %
With the exaltation of computer functions and the rapid development of neural network technology, computer-aided ECG analysis has developed by leaps and bounds. 
However, most works related to the classification of arrhythmia are based on single-lead ECG signals. 
% u should be better than the reference u ref?
% 19_4 patient_specific 12_1 patient_specific
% 18_1 patient_independent
% 21_3 class-oriented
% i am reluctant to cite {12_1} for its bad writtings
Many studies such as \cite{19_4}, \cite{18_1} and \cite{21_3} chose to use the upper lead in the MIT-BIH arrhythmia database (mitdb) but not the lower lead, in which the ventricular ectopic beats (VEB) and supraventricular ectopic beats (SVEB) signals are more conspicuous. 
% cannot compare with ... 
Other studies such as \cite{21_4}, \cite{o_4}, \cite{o_5} and \cite{o_6} only use the lead II ECG signal in their chosen database. 
% ----- multi-lead works ----- %
As far back as 1989, the American Heart Association suggested that monitors should be able to analyze three or more leads \cite{m_11}. 
% m_7 m_8(only used here for a proven)
According to \cite{m_7}, multi-lead ECG analysis is an effective method to improve the detection accuracy and reduce false positive alarms. Yan et al. \cite{m_8} experimentally proved that the use of multi-lead data results in more reliability and higher accuracy.
% o_7(only used here) m_1 m_10 is discarded
However, relatively few studies have used multi-lead ECG data.
In the process of handling multi-lead ECG issues, most works combined the feature vectors by direct concatenation \cite{m_10} and in some cases the final classification report is derived by multi-lead results fusion \cite{m_7}, \cite{m_8}.
Chazal et al. \cite{m_10}, directly connected the feature vectors obtained by sampling each lead of the ECG signal in the feature fusion step.
La et al. \cite{m_1} derive the final prediction by training a set of weighting coefficients for the result produced by each lead.
% m_5(only used here) 
% Shen's work \cite{m_5} use ICA to extract the features of each segment (P wave, QRS interval, ST segment) of one heartbeat of each lead separately.
% the final result is derived by sub-result fusion
The concatenate approach is equivalent to treating the features extracted by different leads as if they are from the ``same lead''. Since the ECG signals are derived from different leads, the positions of the different lead signal acquisition electrodes are different, and they do not have intrinsic homology. The fusion of eigenvectors has no ``lead difference'', and the direct splicing of features is debatable. Automatic feature extraction and feature fusion through neural networks are necessary.

% ----- arrhythmia's lead specificity ----- %
% --- the 2 ref here is too space-taking --- %
It should be noted that some arrhythmias have strong specificities. The QRS waveform characteristics of the left bundle branch block are related to the age of the subject and have lead specificity. \iffalse \cite{15_2}\fi 
% LBBB features(age&lead):
% QRS duration greater than or equal to 120 ms in adults, greater than 100 ms in children 4 to 16 years of age, and greater than 90 ms in children less than 4 years of age. 
% LBBB has very specific waveform characteristics in leads I, avL, V5, V6. R peak time greater than 60 ms in leads V5 and V6 but normal in leads V1, V2, and V3, when small initial R waves can be discerned in the above leads \cite{15_2}.
%%%%%%%% ---- to be confirmed ----- %%%%%%
Most types of arrhythmia diseases such as right bundle branch block and premature beats also conform to the above argument that totally different morphological characteristics appear on different leads. \iffalse \cite{15_1}\fi
% ----- multi lead is good ----- %
Indeed, the 12-lead ECG measurement is cumbersome, and it is practical to use fewer leads to diagnose some of the arrhythmias. 
% evidence for multi-lead is good
The Precise diagnosis of arrhythmias or more complex heart diseases, as well as the localization of lesions, requires a joint analysis of multiple leads. Developing a model that automatically fuses different lead features will increase the upper limit of the ECG analysis algorithm.
% 24_2(only use here) for lesions location prove
% However, combining information from intra-cardiac studies with findings on the 12-lead electrocardiogram (ECG) resulted in much better localization of conduction abnormalities and arrhythmias using the ECG \cite{24_2}.
The greater the number of ECG leads analyzed, the better is the interpretability and reliability of the ``electrophysiological portrait'' of the subject. Therefore, computer-aided multi-lead diagnosis has important research significance and clinical value.

% -- data selected and the established models -- %
Considering the multiple factors above, the mitdb was selected as the research basis of this study. We establish a weight calibration structure called the information-based squeeze and excitation block (ISE-block) for different feature channels in the neural network model to enhance the neural network model's ability to extract features and the ability to achieve feature fusion based on multi-channel ECG input, which can improve the detection and diagnosis level of the CNN model in the ECG signal classification and recognition task. We adopt the end-to-end convolutional neural network model as the backbone of our ISE structure.
% 22_7 is not using in this article!
\iffalse The ECG dataset is processed via the subject-oriented method adopted in \cite{19_4}, \cite{12_1}, \cite{22_4}, \cite{22_7}, \cite{19_6} and \cite{19_7}. \fi
A set of control experiments is conducted to prove that the application of two leads for ECG classification is more effective. For the ISE structure proposed in this study, an ablation study is implemented to verify the benefits of the structure's feature channel recalibration function on the ECG classification task. We introduce patient-specific scheme for the comparison with some state-of-the-art models, in which an individual attention model is trained for each patient.

% ----- paper structure prefix ----- %
% -- may need to rewrite the rhetoric of this -- %
The rest of this article is organized as follows. Section II outlines the ECG dataset used in this study, as well as the corresponding dataset preprocessing methods for the control trials, including descriptions of the lead configuration, ECG data processing and dataset partitioning. Several network models for the ablation experiments are presented in Section III, and the structure of the proposed ISE-block, together with corresponding basic mathematical expressions is described. In Section IV, we introduce the objective function, the algorithm optimizer, the evaluation index of the proposed model, and the framework used for programming. The results of all the control experiments performed in this study are presented in Section V with careful analysis. Section VI summarizes the experimental results and provides a profound discussion of the principles and performance of the proposed structure. Section VII summarizes the work of the entire study. The theoretical basis for supporting ISE-block design is neatly expressed in Appendix A.

% ----- Methodology ----- %
\section{Methodology}
\subsection{Database}
% ----- database ----- %
% ----- citation range should be re-check ----- %
The mitdb \cite{2} is used in this study. The benchmark database was established by the Beth-Israel Hospital in Boston and the MIT Lab. It is the first dataset to contain standard test materials that are widely used to evaluate the performance of arrhythmia detectors, and is the only one indicated by the Association for the Advancement of Medical Instrumentation (AAMI) standard that contemplates all 5 of the superclasses for arrhythmias \cite{o_8}. This database has been widely used in some remarkable studies \cite{19_1}, \cite{19_4}, \cite{22_7}, \cite{19_7}, \cite{19_6}, and \cite{22_4}. The database contains 48 thirty minute-long two-channel Holter recordings obtained from 47 subjects in the BIH Arrhythmia Laboratory from 1975 to 1979. Twenty-three of the records are randomly selected from a dataset containing 4000 twenty-four-hour dynamic ECG records, and the other 25 records are selected from the same dataset to contain some uncommon but clinically significant cases. The ECG data of each channel in the database are digitized at a specification of 360 Hz sampling points per second. Each beat is labeled by two or more cardiologists for each record in the database, with the R-wave peak or local extrema of the beat noted. \iffalse The statistics of the ECG samples of each record in the MIT-BIH arrhythmia database are shown in Table \ref{table0}.\fi

% ----- signal preprocessing ----- %
% implement different filtering method for the upper-lead and lower lead
% 
\subsection{ECG Signal Preprocessing}
The mitdb contains more than 30 kinds of annotations, including beat-level annotations, wave-level annotations, and signal-related annotations, in which a variety of arrhythmia signals are attentively annotated. Although various types of cardiac arrhythmias exist, AAMI \cite{6_1} recommends that only some types should be detected by equipment or algorithms \cite{o_8}. According to the AAMI recommendations, there are 17 recommended arrhythmia categories that are classified into 5 superclasses. \iffalse more details about this mapping rules see Table. \ref{table0}\fi The following mappings of categories are based on \cite{19_2}: N: Normal beat (1), Left bundle branch block beat (2), Right bundle branch block beat (3), and Left or right bundle branch block (25); S: Aberrated atrial premature beat (4), Nodal (junctional) premature beat (7), Atrial premature contraction (8), Premature or ectopic supraventricular beat (9), Nodal (junctional) escape beat(11), Atrial escape beat (34), and Supraventricular escape beat (35); V: Premature ventricular contraction (5), and Ventricular escape beat (10); F: Fusion of ventricular and normal beat (6); and Q: Paced beat (12), Unclassifiable beat (13), and Fusion of paced and normal beat (38). All data used in the experiments in this study were processed according to the classification suggested by the practice recommended by AAMI ECAR-1987 \cite{6_1}.  % (used 2 times)

\begin{figure}[thbp]
\centering
\includegraphics[width=3.4in]{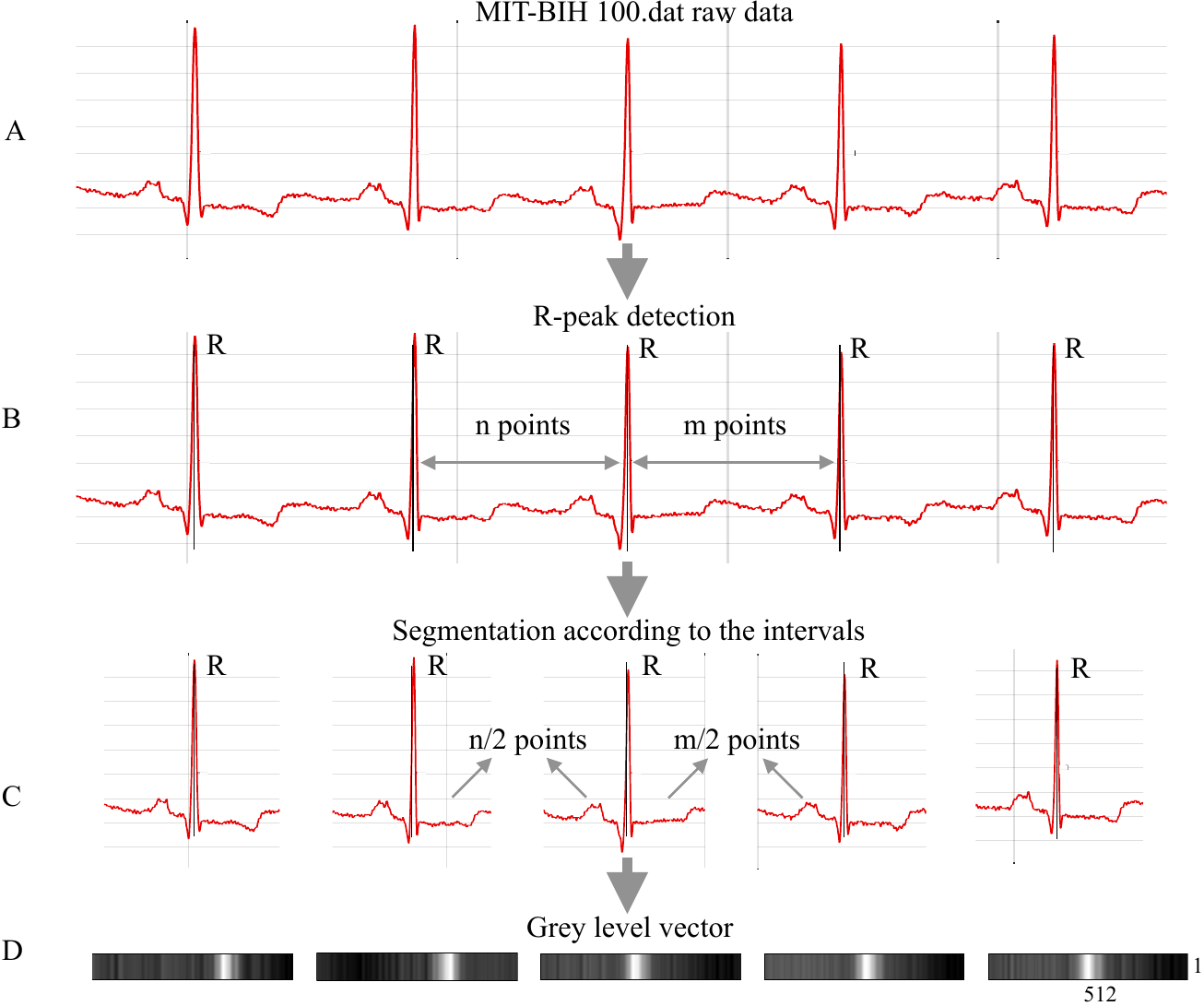}
\caption{Data preprocessing pipeline using heartbeats in the mitdb record 100.}
\label{figure1}
\vspace{-5pt}
\end{figure}
% ----- lead configuration ----- %
\subsubsection{Lead Configuration}In most records of the mitdb, the upper signal is a modified limb lead II (MLII), obtained by placing the electrodes on the chest. The lower signal is usually a modified lead V1 (occasionally V2 or V5, and in one instance, V4), and as for the upper signal, the electrodes are also placed on the chest \cite{4} \cite{5}. According to the website located at \url{https://www.physionet.org}, normal QRS complexes are usually prominent in the upper signal, while the normal beats are frequently difficult to discern in the lower signal, although ectopic beats (SVEB and VEB) will often be more prominent. Considering the above factors, three lead configurations are adopted for the control experiments: ML\RNum{2}, V1, and ML\RNum{2} together with V1.

% ----- record excluded ----- %
\subsubsection{Record Excluded}
In records 102 and 104, it was not possible to use the modified lead II because of surgical dressings on the patients. Four of the 48 records (102, 104, 107, and 217) include paced beats \cite{4} \cite{5}, so we exclude them in our experiments. Since we need to use two lead signals, we must ensure the consistency of the leads in all the samples---the first lead must be a modified lead II and the second lead is V1. Records 100, 103, 114, 117, 123 and 124 are excluded because their second signal was not acquired via lead V1. The remaining 38 records are used as the dataset for the following experiments.

% ----- data refactor ----- %
% ----- the footnote can be deleted ----- %
\subsubsection{Data Refactoring Pipeline}
\begin{figure}[htbp]
\vspace{0pt}
\setlength{\belowcaptionskip}{0pt}{
\centering
\includegraphics[width=3.5in]{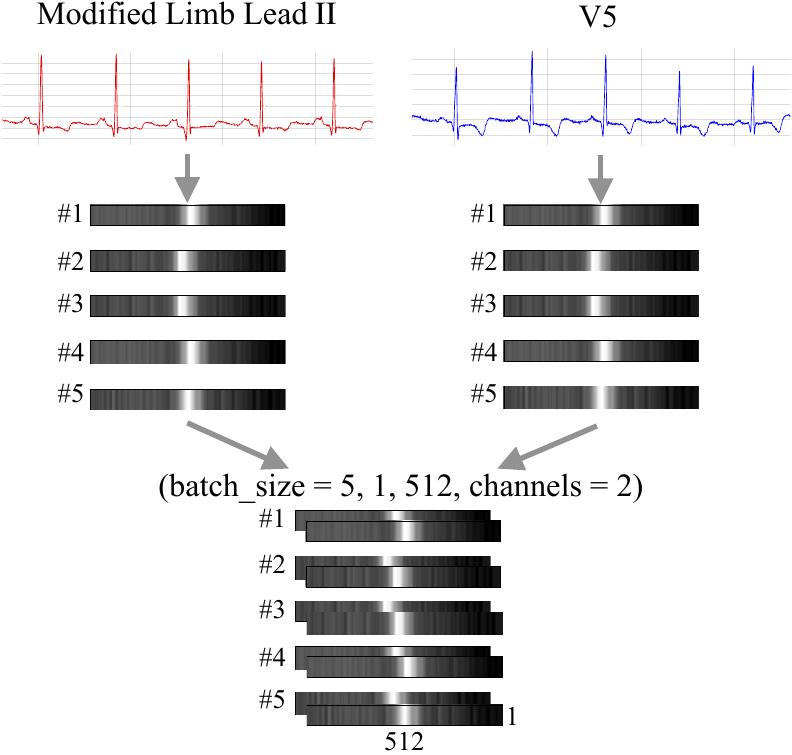}
\caption{Reconstruction of two-lead ECG data using MIT-BIH Record 100 as a case. The output data is organized in the form of $(batch\_size=5, height=1, width=512, channel=2)$.}
\label{figure2}}
\vspace{-10pt}
\end{figure}

Fig. \ref{figure1} shows the pipeline of the ECG data preprocessing procedures. We split the ECG signal in a similar way to that used in \cite{18_1}, but after the segmentation, the alignment step is not used. Part A in Fig. \ref{figure1} shows the original ECG signal in the mitdb converted from the format 212 to the analog electrical signal\footnote{More details about format 212 can be found at: https://physionet.org/physi\\otools/wag/signal-5.htm\#sect9}.
% \url{https://physionet.org/physiotools/wag/signal-5.htm#sect9}
% ----- R-peak detection ----- %
% 10_1 10_4 10_5 only cited once
R-peak detection is shown in part B in Fig. \ref{figure1}. Many advanced R-peak detection methods such as \cite{10_1}, \cite{10_4}, and \cite{10_5} can be used for the mitdb. We directly split the database according to its original annotation.
% ----- operation of segmentation ----- %
The time series of ECG signals are separated according to the sampling rate and the locations of the R-peaks, as shown in part C in Fig. \ref{figure1}. Specifically, the signal is cut at beat by beat at the points, which is half of the number the sampling points away from two adjacent R peaks, to produces dynamically segmented heartbeats. The number of sampling points for one segmented beat is computed as follows:
% ----- operation of segmentation ----- %
\begin{equation}
\begin{split}
N_{R_j}= I_{R_j}+I_{R_{j+1}} + 1
\end{split}
\end{equation}
$R_j$ is the $j$-th R peak, $I_{R_j}$ is the number of sampling points of the interval of $R_j$ and $R_{j-1}$, and $N_{R_j}$ is the number of sampling points of $j$-th ECG waveform centered on the $j$-th R-peak. \iffalse It should be noted that this ECG splitting method will not work on the first and last ECG waveforms in each record. The number of waveforms available in each record will be 2 less than the number of the ECG heartbeats in each record.\fi
% ----- grey scale them for display----- %
% the filter configuration need to be in details?
Segmented ECG heartbeats are transformed into 1D vectors, as shown in part D in Fig. \ref{figure1}. \iffalse The floats are transformed to the nearest integers, which can speed up the calculation of the model but loses a little accuracy.\fi We implement necessary filtering of the signal to remove high-frequency noise and baseline drift and the normalization of the analog voltage values in each beat is adopted to accelerate the training convergence process.
% ----- interpolation or decimate ----- %
To best preserve the morphological features of the heartbeat samples, while taking into account the model's algorithm and implementation framework, statistical analysis is used to overview the length information of the samples in the training set and the cubic spline interpolation method is used to supplement the heartbeats in a batch to the same length of 512 points. Some training heartbeats segmented has irregular length, we remove them for better performance of our algorithm. Random downsampling is used for the samples above 512 points. The way data are preprocessed, especially for signals with timing characteristics, is of great importance, and inappropriate processing will result in classification mess. For premature beat waveforms, padding zeros in front of the waveform may cause serious characteristic changes, and it can be the reason why many proposed algorithms are confused in distinguishing between N and S classes.
% ----- reconstruction ----- %
The ECG signals of the two leads in the mitdb used for models that accept dual-lead signal inputs are reconstructed in the manner shown in Fig. \ref{figure2}. \iffalse The digital voltage values of the beats are normalized into a scale of 0-255 to better display of the ECG sample reconstruction process.\fi

\subsubsection{Separate Training Test}
% ye's work is on patient-independent scheme.
According to the work of \cite{20_2}, there are two ways to divide the dataset, which are called ``class-oriented'' and ``subject-oriented'' methods. We implemented the subject-oriented data partitioning method to conduct a series of control experiments and the patient-specific scheme is taken into consideration.
\iffalse The relevant data partitioning methods are used to verify the beneficial properties of our proposed neural network structure compared with the plain Resnet backbone structure and they are used to support the view that utilizing multi-lead data in heartbeat classification modeling enhances the performance of the deep networks. The subject-oriented test set is used to demonstrate the generalization performance of our proposed model in the MIT-BIH arrhythmia database.\fi
% ----- data reconstruction ----- %
%We divide 38 records into two sub-datasets: DS1 and DS2. DS1 contains data from ECG recordings: 101, 105, 106, 108, 109, 111, 112, 113, 115, 116, 118, 119, 121 and 122; DS2 contains data from ECG recordings: 200, 201, 202, 203, 205, 207, 208, 209, 210, 212, 213, 214, 215, 219, 220, 221, 222, 223, 228, 230, 231, 232, 233 and 234. According to the evaluation scheme recommended by AAMI, the VEB and SVEB can be evaluated separately using the two subsets: DS2-v and DS2-s. DS2-v contains data from: 200, 202, 210, 213, 214, 219, 221, 228, 231, 233 and 234; DS2-s contains data from: 200, 202, 210, 212, 213, 214, 219, 221, 222, 228, 231, 232, 233 and 234.
We divide 38 records into two sub-datasets\footnote{DS1 contains data from ECG recordings: 101, 105, 106, 108, 109, 111, 112, 113, 115, 116, 118, 119, 121 and 122; DS2 contains data from ECG recordings: 200, 201, 202, 203, 205, 207, 208, 209, 210, 212, 213, 214, 215, 219, 220, 221, 222, 223, 228, 230, 231, 232, 233 and 234.}: DS1 and DS2. According to the evaluation scheme recommended by AAMI, the VEB and SVEB can be evaluated separately using the two subsets: DS2-v and DS2-s\footnote{DS2-v contains data from: 200, 202, 210, 213, 214, 219, 221, 228, 231, 233 and 234; DS2-s contains data from: 200, 202, 210, 212, 213, 214, 219, 221, 222, 228, 231, 232, 233 and 234.}.
The first 5 minutes of the ECG records (DS1 and DS2) are used as the training set, which builds up the common part and the patient-specific part, and the remaining 25 minutes of each segment are used as test set. The numbers of samples in AAMI classes for the training set are 13480 ($N$), 258 ($S$), 1083 ($V$), 175 ($F$), 0 ($Q$). The numbers are 65524 ($N$), 2668 ($S$), 5828 ($V$), 618 ($F$), 13 ($Q$) for the test set. \iffalse A few heartbeats in training sets with morphological anomalies due to the segmentation algorithm proposed above are manually deleted.\fi

% ----- data augmentation ----- %
% borderline-smote citation
% H. Han, W. Wen-Yuan, M. Bing-Huan, "Borderline-SMOTE: a new over-sampling method in imbalanced datasets learning," Advances in intelligent computing, 878-887, 2005.
% 13_1's paper is not found yet...
\subsubsection{Data Equilibrium}
To prevent unbalanced samples from causing an incorrect propensity when training the model, we applied the method of label shuffling method \cite{13_1} to augment the data for each training set. Simply applying the duplication to upsample the lack-of-sample classes will not increase the amount of substantial information, so we use the SMOTE method \cite{smote} to apply data augmentation to the lack-of-sample classes (SVEB) and edit the Tomek links \cite{tomek} for all the training set, mainly to reduce unnecessary N-class samples, enhancing the generalization ability of the trained network model to find a balanced performance for the easy-to-confuse N and S classes. In addition, adopting both duplication and downsampling can lighten the gravity of the overfitting problem. 
% the numbers are to be check...!!!!!!!
After the augmentation, the final number of samples used for training after balance is as follows:  10000 ($N$), 8000 ($S$), 8000 ($V$), 4000 ($F$), 0 ($Q$). The class Q, excluded in augmentation procedures due to its extremely small number of samples, is not consider as main index of the evaluation.

% ----- TABLE -1(net structure) ----- %
% output size is not defined yet!!!
\begin{table*}[htbp]
%\begin{center}
\centering
\caption{Backbone Network Structures with Various Depth for Model Complexity Mining towards ECG Classification Task}
% (the output size should be defined by debug)
\label{table1}
\setlength{\arraycolsep}{1.8pt}
\setlength{\tabcolsep}{0.1mm}{
\begin{tabular}{c|c|cccccc}
\hline
Layer & output & \multicolumn{1}{c|}{11-layer} & \multicolumn{1}{c|}{14-layer} & \multicolumn{1}{c|}{17-layer} & \multicolumn{1}{c|}{20-layer} & \multicolumn{1}{c|}{23-layer} & 26-layer \\ 
\hline
conv1 & \,$1\times256$\, & \multicolumn{5}{c}{1x4, 8, stride 2} \\ 
\hline
\multirow{4}{*}{ block1 } & \multirow{4}{*}{$1\times64$} & \multicolumn{5}{c}{1x3 max pool, stride 2} \\ 
\cline{3-8} 
%& & \multicolumn{1}{c|}{\multirow{3}{*}{\begin{tabular}[c]{@{}c@{}} a \\ b \\ c \\ \end{tabular}}}  % 1
& & \multicolumn{1}{c|}
{\multirow{3}{*}{
\renewcommand{\arraystretch}{0.8}  % set as 0.8
$
\left[
\begin{array}{cccc} 
    1\times1, & \ 8 \\ 
    1\times3, & \ 8 \\
    1\times1, & \ 16 \\  
\end{array}
\right]
\times 1\ $}}
& \multicolumn{1}{c|}
{\multirow{3}{*}{
\renewcommand{\arraystretch}{0.8}  % set as 0.8
$
\left[
\begin{array}{cccc} 
    1\times1, & \ 8 \\ 
    1\times3, & \ 8 \\
    1\times1, & \ 16 \\  
\end{array}
\right]
\times 1\ $}}
& \multicolumn{1}{c|}{
\multirow{3}{*}{
\renewcommand{\arraystretch}{0.8}  % set as 0.8
$
\left[
\begin{array}{cccc} 
    1\times1, & \ 8 \\ 
    1\times3, & \ 8 \\
    1\times1, & \ 16 \\  
\end{array}
\right]
\times 1\ $}} 
& \multicolumn{1}{c|}{
\multirow{3}{*}{
\renewcommand{\arraystretch}{0.8}  % set as 0.8
$
\left[
\begin{array}{cccc} 
    1\times1, & \ 8 \\ 
    1\times3, & \ 8 \\
    1\times1, & \ 16 \\  
\end{array}
\right]
\times 1\ $}}
& \multicolumn{1}{c|}{
\multirow{3}{*}{
\renewcommand{\arraystretch}{0.8}  % set as 0.8
$
\left[
\begin{array}{cccc} 
    1\times1, & \ 8 \\ 
    1\times3, & \ 8 \\
    1\times1, & \ 16 \\  
\end{array}
\right]
\times 1\ $}}
& \multicolumn{1}{c}{
\multirow{3}{*}{
\renewcommand{\arraystretch}{0.8}  % set as 0.8
$
\left[
\begin{array}{cccc} 
    1\times1, & \ 8 \\ 
    1\times3, & \ 8 \\
    1\times1, & \ 16 \\  
\end{array}
\right]
\times 1\ $}} \\
& & \multicolumn{1}{c|}{} & \multicolumn{1}{c|}{} & \multicolumn{1}{c|}{} & \multicolumn{1}{c|}{} & \multicolumn{1}{c|}{} \\
& & \multicolumn{1}{c|}{} & \multicolumn{1}{c|}{} & \multicolumn{1}{c|}{} & \multicolumn{1}{c|}{} & \multicolumn{1}{c|}{} \\ 
\hline
\multirow{3}{*}{ block2 } & \multirow{3}{*}{$1\times32$} 
%& \multicolumn{1}{c|}{\multirow{3}{*}{\begin{tabular}[c]{@{}c@{}} e \\ f \\ g \\ \end{tabular}}}  % 2
& \multicolumn{1}{c|}{
\multirow{3}{*}{
\renewcommand{\arraystretch}{0.8}  % set as 0.8
$
\left[
\begin{array}{cccc} 
    1\times1, & \ 16 \\ 
    1\times3, & \ 16 \\
    1\times1, & \ 32 \\  
\end{array}
\right]
\times 1\ $}}
& \multicolumn{1}{c|}{
\multirow{3}{*}{
\renewcommand{\arraystretch}{0.8}  % set as 0.8
$
\left[
\begin{array}{cccc} 
    1\times1, & \ 16 \\ 
    1\times3, & \ 16 \\
    1\times1, & \ 32 \\  
\end{array}
\right]
\times 1\ $}}
& \multicolumn{1}{c|}{
\multirow{3}{*}{
\renewcommand{\arraystretch}{0.8}  % set as 0.8
$
\left[
\begin{array}{cccc} 
    1\times1, & \ 16 \\ 
    1\times3, & \ 16 \\
    1\times1, & \ 32 \\  
\end{array}
\right]
\times 1\ $}} 
& \multicolumn{1}{c|}{
\multirow{3}{*}{
\renewcommand{\arraystretch}{0.8}  % set as 0.8
$
\left[
\begin{array}{cccc} 
    1\times1, & \ 16 \\ 
    1\times3, & \ 16 \\
    1\times1, & \ 32 \\  
\end{array}
\right]
\times 2\ $}}
& \multicolumn{1}{c|}{
\multirow{3}{*}{
\renewcommand{\arraystretch}{0.8}  % set as 0.8
$
\left[
\begin{array}{cccc} 
    1\times1, & \ 16 \\ 
    1\times3, & \ 16 \\
    1\times1, & \ 32 \\  
\end{array}
\right]
\times 2\ $}} 
& \multicolumn{1}{c}{
\multirow{3}{*}{
\renewcommand{\arraystretch}{0.8}  % set as 0.8
$
\left[
\begin{array}{cccc} 
    1\times1, & \ 16 \\ 
    1\times3, & \ 16 \\
    1\times1, & \ 32 \\  
\end{array}
\right]
\times 2\ $}} \\ 
& & \multicolumn{1}{c|}{} & \multicolumn{1}{c|}{} & \multicolumn{1}{c|}{} & \multicolumn{1}{c|}{} & \multicolumn{1}{c|}{}\\
& & \multicolumn{1}{c|}{} & \multicolumn{1}{c|}{} & \multicolumn{1}{c|}{} & \multicolumn{1}{c|}{} & \multicolumn{1}{c|}{}\\ 
\hline
\multirow{3}{*}{ block3 } & \multirow{3}{*}{$1\times16$} 
%& \multicolumn{1}{c|}{\multirow{3}{*}{\begin{tabular}[c]{@{}c@{}} h \\ i \\ j \\ \end{tabular}}}   % 3
& \multicolumn{1}{c|}{
\multirow{3}{*}{
\renewcommand{\arraystretch}{0.8}  % set as 0.8
$
\left[
\begin{array}{cccc} 
    1\times1, & \ 32 \\ 
    1\times3, & \ 32 \\
    1\times1, & \ 64 \\  
\end{array}
\right]
\times 1\ $}}
& \multicolumn{1}{c|}{
\multirow{3}{*}{
\renewcommand{\arraystretch}{0.8}  % set as 0.8
$
\left[
\begin{array}{cccc} 
    1\times1, & \ 32 \\ 
    1\times3, & \ 32 \\
    1\times1, & \ 64 \\  
\end{array}
\right]
\times 1\ $}}
& \multicolumn{1}{c|}{
\multirow{3}{*}{
\renewcommand{\arraystretch}{0.8}  % set as 0.8
$
\left[
\begin{array}{cccc} 
    1\times1, & \ 32 \\ 
    1\times3, & \ 32 \\
    1\times1, & \ 64 \\  
\end{array}
\right]
\times 2\ $}} 
& \multicolumn{1}{c|}{
\multirow{3}{*}{
\renewcommand{\arraystretch}{0.8}  % set as 0.8
$
\left[
\begin{array}{cccc} 
    1\times1, & \ 32 \\ 
    1\times3, & \ 32 \\
    1\times1, & \ 64 \\  
\end{array}
\right]
\times 2\ $}}
& \multicolumn{1}{c|}{
\multirow{3}{*}{
\renewcommand{\arraystretch}{0.8}  % set as 0.8
$
\left[
\begin{array}{cccc} 
    1\times1, & \ 32 \\ 
    1\times3, & \ 32 \\
    1\times1, & \ 64 \\  
\end{array}
\right]
\times 2\ $}}
& \multicolumn{1}{c}{
\multirow{3}{*}{
\renewcommand{\arraystretch}{0.8}  % set as 0.8
$
\left[
\begin{array}{cccc} 
    1\times1, & \ 32 \\ 
    1\times3, & \ 32 \\
    1\times1, & \ 64 \\  
\end{array}
\right]
\times 3\ $}} \\
& & \multicolumn{1}{c|}{} & \multicolumn{1}{c|}{} & \multicolumn{1}{c|}{} & \multicolumn{1}{c|}{} & \multicolumn{1}{c|}{}\\
& & \multicolumn{1}{c|}{} & \multicolumn{1}{c|}{} & \multicolumn{1}{c|}{} & \multicolumn{1}{c|}{} & \multicolumn{1}{c|}{}\\ 
\hline
\multirow{3}{*}{ block4 } & \multirow{3}{*}{$1\times8$} 
%& \multicolumn{1}{c|}{\multirow{3}{*}{\begin{tabular}[c]{@{}c@{}} k \\ l \\ m \\ \end{tabular}}}  % 4
& \multicolumn{1}{c|}{
\multirow{3}{*}{
\renewcommand{\arraystretch}{0.8}  % set as 0.8
%$
%\left[
%\begin{array}{cccc} 
%    1\times1, & 64 \\ 
%    1\times3, & 64 \\
%    1\times1, & 128 \\  
%\end{array}
%\right]
%\times 1\ $
}}
& \multicolumn{1}{c|}{
\multirow{3}{*}{
\renewcommand{\arraystretch}{0.8}  % set as 0.8
$
\left[
\begin{array}{cccc} 
    1\times1, & 64 \\ 
    1\times3, & 64 \\
    1\times1, & 128 \\  
\end{array}
\right]
\times 1\ $}
}
& \multicolumn{1}{c|}{
\multirow{3}{*}{
\renewcommand{\arraystretch}{0.8}  % set as 0.8
$
\left[
\begin{array}{cccc} 
    1\times1, & 64 \\ 
    1\times3, & 64 \\
    1\times1, & 128 \\  
\end{array}
\right]
\times 1\ $}} 
& \multicolumn{1}{c|}{
\multirow{3}{*}{
\renewcommand{\arraystretch}{0.8}  % set as 0.8
$
\left[
\begin{array}{cccc} 
    1\times1, & 64 \\ 
    1\times3, & 64 \\
    1\times1, & 128 \\  
\end{array}
\right]
\times 1\ $}}
& \multicolumn{1}{c|}{
\multirow{3}{*}{
\renewcommand{\arraystretch}{0.8}  % set as 0.8
$
\left[
\begin{array}{cccc} 
    1\times1, & 64 \\ 
    1\times3, & 64 \\
    1\times1, & 128 \\  
\end{array}
\right]
\times 2\ $}} 
& \multicolumn{1}{c}{
\multirow{3}{*}{
\renewcommand{\arraystretch}{0.8}  % set as 0.8
$
\left[
\begin{array}{cccc} 
    1\times1, & 64 \\ 
    1\times3, & 64 \\
    1\times1, & 128 \\  
\end{array}
\right]
\times 2\ $}} \\
& & \multicolumn{1}{c|}{} & \multicolumn{1}{c|}{} & \multicolumn{1}{c|}{} & \multicolumn{1}{c|}{} & \multicolumn{1}{c|}{}\\
& & \multicolumn{1}{c|}{} & \multicolumn{1}{c|}{} & \multicolumn{1}{c|}{} & \multicolumn{1}{c|}{} & \multicolumn{1}{c|}{}\\ 
\hline
& $1\times1$ & \multicolumn{5}{c}{global average pooling} \\ 
\hline
\end{tabular}}
\end{table*}

\section{Models}
\subsubsection{Resnet for a single channel --- model A}
This model is built with a single channel of ECG input. We performed experiments on each lead of the signal. 
% delete for no use sentence
\iffalse The experimental results validate the MIT-BIH database's description of the differences in the significance between normal and abnormal waveforms in different leads.\fi
We adopt the residual-v2 blocks \cite{14_3} structure to build the ECG signal classification network model. The residual learning framework structure is a network with a shortcut, which solves the issue of training a deeper neural network model being difficult. We build a series of deep neural networks with residual blocks. See Table \ref{table1} for details of the models.
\subsubsection{Resnet for both channels --- model B}
We build a two-channel ECG inputs model, which takes the advantage of the information gain by increasing the feature dimensions of the input that is organized in the form shown in Fig. \ref{figure2}. In addition, we attempt to establish relationships between different channels (integrating information from different channels) through the architecture of the plain residual CNN. The backbone structure of this model is the same as that of model A shown in Table \ref{table1}.
% ----- ise-block ----- %
\begin{figure}[htbp]
\centering
\includegraphics[width=3.5in]{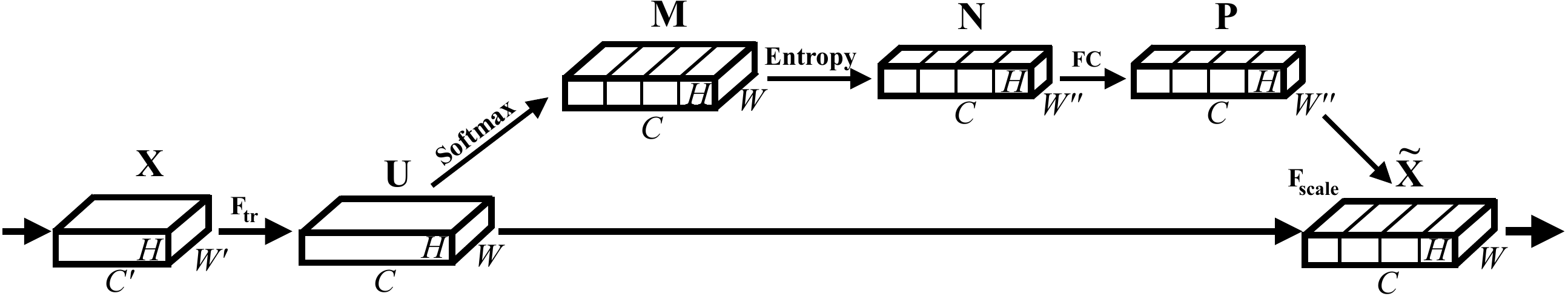}
\caption{Information-based squeeze and excitation block. $H$ stands for the height of the feature map. $W$ is the width of the feature map. $W'$ and $W''$ indicates a different numerical value from $W$. And $C$ is the number of channels of the feature map. Bold characters $XUMNP$ are names of the matrices. $F_{tr}$ is the residual block in the proposed model and $F_{scale}$ is the element-wise multiply with the broadcast nature in Python between the P and U matrices.}
\label{figure3}
\vspace{-5pt}
\end{figure}

\subsubsection{Attention model for both channels --- model C}
% simple depict the model-3 with ISE
Compared with model B, the major improvement of this model is that a channel-attention network structure named as ISE-block, is embedded to realize channel-wise feature maps recalibration. The backbone structure of this model is the same as those of model A and model B shown in Table \ref{table1}.
% ----- softmax figure ----- %
\begin{figure}[bp]
\centering
\includegraphics[width=3.5in]{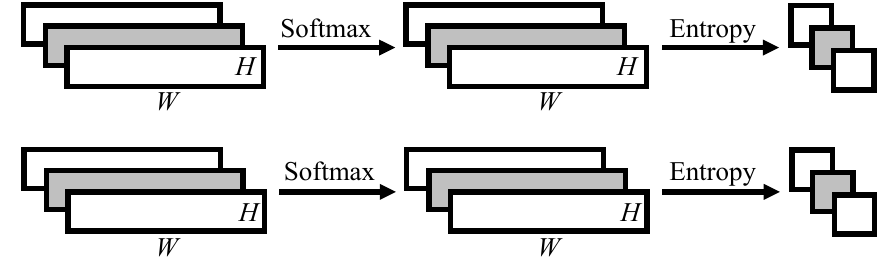}
\caption{Softmax and Entropy block. A brief schematic of the softmax and entropy implementation of the data batch in form of $(batch\_size=2, height=H, width=W, channel=3)$.}
\label{figure4}
\end{figure}
% Macro introduction to the proposed structure %
An ISE-block is established on the output feature maps of different channels, and it attempts to describe the difference in the importance and richness of valid information in different feature maps of the same hidden layer and attempts to weight the features accordingly. For the task of multi-channel ECG classification, we argue that channel-wise learning can be used to improve the model's ability to express the multi-lead ECG data. \iffalse It should be emphasized the channels of a channel-wise learning algorithm and of a multi-channel ECG signal has different meanings. Each time the convolution of the input feature maps is performed, the fusion of the information in multi-channel input is involved.\fi Compared with Hu's work \cite{14_2}, we build a more complex channel-attention structure, through which the information-based feature calibration is realized. 
\iffalse We apply the structures of softmax and entropy to achieve global information embedding for the feature maps generated in the middle of deep neural networks.\fi
The following describes the details of this channel-wise structure.

% ----- define the feature map by math ----- %
According to our network definition, the data format of the feature maps can be expressed in the matrix form of $(batch\_size, height, width, channel)$. For a clear display of the ISE-block, let $batch\_size=1$, and then the ISE-block is as shown in Fig. \ref{figure3}. It is possible to refer to the feature maps obtained by the forward propagation of all the ECG samples in the input batch corresponding to the layer $l$ of the convolutional neural network model as $F_l$. Therefore, $F_l[i],i=0\dots(batch\_size-1)$ represents a set of feature maps which are obtained by the forward propagation of the $i$-th ECG sample in an input batch to the $l$-th layer in the neural network model, which has a format of $(height, width, channel)$.
As shown in Fig. \ref{figure4}, we implement the softmax function to transform the feature map U to feature map M in Fig. \ref{figure3}. Given the general implementation of softmax formula as follows:
% ----- softmax equation----- %
\begin{equation}
\begin{split}
\sigma(\bm{z_j})=\frac{e^{z_j}}{\sum_{k=1}^Ke^{z_k}}
\end{split}
\end{equation}
We build the softmax structures for the feature map derived by the forward propagation of the input batch. The feature maps taken into consideration for the softmax computation can be defined as follows: Using the $F_l[i]$ defined above as the base, the $c$-th channel of the set of feature maps is defined as $F_l[i][c]$, and the value of the $j$-th node in $F_l[i][c]$ is set as $F_l[i][c][j]$. Thus, the softmax computation can be presented by the following equation:
% ----- feature map definition for softmax ----- %
\begin{equation}
\begin{split}
\sigma(F_l[i][c][j])=\frac{e^{F_l[i][c][j]}}{\sum_{j=0}^{W\times H-1}e^{F_l[i][c][j]}}
\end{split}
\end{equation}
where the $\sigma$ is the symbolic representation of the softmax function. 
\iffalse Given the equation above, we attempt to calculate the true distribution of the feature map by calculating the softmax function of a batch input sample in the specified feature map.\fi
% ----- about entropy ----- layer %
The entropy structure is shown in the M to N section of Fig. \ref{figure3}, which is named as ``Extraction''. As shown in the Fig. \ref{figure4}, the softmax layers proposed above do not change the data format of the feature map in the convolutional neural network. Given the symbol $F_l[i][c][j]$ as above, the entropy is computed as: 
% ----- entropy ----- %
\begin{equation}
\begin{split}
H(\bm{z})=-\sum_{j=0}^{W\times H-1} p(F_l[i][c][j])\log p(F_l[i][c][j])\label{e4}
\end{split}
\end{equation}
where $z$ stands for the random variable of the $c$-th channel of the $l$-th hidden layer.
% ----- follow up ----- %
It should be noted that the above formula is obtained on the premise that several approximations are applied and will be discussed in detail in the appendixes.
% ----- FC layer ----- %
\begin{figure}[htbp]
\centering
\includegraphics[width=1.5in]{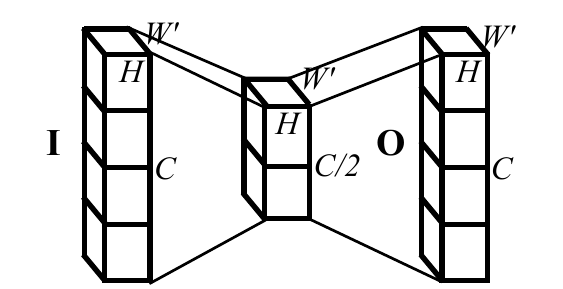}
\caption{Full-connected layers with decay ratio=2.}
\label{figure5}
\end{figure}
The fully-connected layers (FC layers) work as a ``weight calibrator'' for the information embedded vector shown in Fig. \ref{figure5}. In the design of this structure, the initial motivation is that the training of two FC layers can compensate for the inherent shortcomings of the structure N's deficiency in mapping its values to the relative valid information quantity of $F_l[i][c]$. We also refer to the two standards proposed in the SE-net \cite{14_2}: First, the structure should be capable of learning a nonlinear interaction between channels and second, it must learn a non-mutually-exclusive relationship. The number of input nodes in the FC layers is equal to the number of its output nodes, which is equal to the number of channels in the ISE-block input feature map. Assuming that the number of channels in the ISE-block's input feature map is $\bm{c}$, the ratio ($\bm{r}$) represents the decay rate of the channels input, then the number of hidden nodes ($\bm{n}$) is determined by the following formula:
% ----- num of units in hidden layer ----- %
\begin{equation}
\begin{split}
& \bm{n}=\lfloor \bm{c/r} \rfloor \\
& \bm{r}=2 , \bm{c}<=8\quad or \quad \bm{r}=8, \bm{c}>8. \label{e5}
\end{split}
\end{equation}
where $\lfloor a \rfloor$ stands for the integer ($floor$) of a.
Finally, we use the structure named P in Fig. \ref{figure3} as a fuzzy measure of the relative validity of the information extracted from each feature map in a certain layer.

% ------- training ------ %
\section{Training}
\subsubsection{Training objective}
The open source framework TensorFlow \iffalse \cite{tensorflow}\fi is adopted for all the experiments conducted, and the $softmax\_cross\_entropy\_with\_logits\_v2$ function in TensorFlow is utilized to calculate the cross entropy. The l2-loss for the weights in the network is added to the total loss to resist the overfitting problem.
\subsubsection{Optimizer selection}The adam optimizer and the momentum optimizer are considered in our experiments. The adam optimizer always performs better than the momentum optimizer in the training process, and it produces good test results in most cases. In general, models optimized with momentum have stronger generalization capabilities in ECG classification issues in testing procedures. The optimizer related parameters are set as follows. For the adam optimizer: beta1: $0.9$, beta2: $0.999$, and epsilon: $1\times e^{-8}$. For the momentum optimizer: momentum: $0.9$, and use\_nesterov=False.
\subsubsection{Learning rate scheme}We tested two methods for decreasing the learning rate---the piecewise constant learning rate and exponential decay learning rate, and found that the exponential decay learning rate  can achieve better results in the training set, yet the piecewise-constant learning always yields better results with carefully designed learning rates and learning rate boundaries in testing procedures. The relevant parameters are set as follows: For exponential-decay, decay steps: $1$ epoch, and decay rate: $0.97$; For the piecewise constant, learning rate list: $1\times 10^{-3}$, $5\times 10^{-4}$, $1\times 10^{-4}$, $5\times 10^{-5}$, $1\times 10^{-5}$, $5\times 10^{-6}$, and $1\times 10^{-6}$ and the learning rate boundaries: $3$-epochs, $10$-epochs, $20$-epochs, $40$-epochs, $80$-epochs, and $160$-epochs. The hyper-parameters listed above are set as a base for most of the ablation experiments. Fine-tuning is necessary for different backbone selections.
\subsubsection{Training strategy}
All the models in this article are trained with Nvidia Tesla P100 GPU, and the code is written in Python 3.6.5.
In the subject-oriented experiments, the method of cross-validation on the training and validation sets is used to obtain relatively reliable hyper-parameters and to obtain a more generalized model that does not depend on the quality of data partitioning.
In addition to the above strategies, we introduce the ``early stop'' method to explore the best performance to the intractable problem of misclassifying S into N and finally reach a balance between the chosen indicators: sensitivity and precision. Specifically, we perform model evaluation on the validation set per epoch training steps. The performance of each model on the dataset per epoch is recorded by our program, and the checkpoints with better performances are saved automatically.
\subsubsection{Assessment indicators}
We evaluated the results obtained from the experiments according to the evaluation indicators recommended in \cite{6_2} and with reference to the evaluation methods in other papers for ECG classification in the mitdb. The final evaluation indicators are as follows:
% ----- assessment indicators ----- %
% acc
\begin{equation}
\begin{split}
Acc=\frac{TP+TN}{TP+TN+FP+FN}
\end{split}
\end{equation}
% sen
\begin{equation}
\begin{split}
Sen=\frac{TP}{TP+FN}
\end{split}
\end{equation}
% spe
\begin{equation}
\begin{split}
Spe=\frac{TN}{TN+FP}
\end{split}
\end{equation}
% ppr
\begin{equation}
\begin{split}
Ppr=\frac{TP}{TP+FP}
\end{split}
\end{equation}
% mcc
\begin{equation}
\begin{split}
MCC=\frac{TP/N-S\times P}{\sqrt{PS(1-S)(1-P)}}
\end{split}
\end{equation}
where TP is true positive, TN is true negative, FP is false positive and FN is false negative, MCC is the Matthews correlation coefficient, $N=TN+TP+FN+FP$, $S=(TP+FN)/N$, and $P=(TP+FP)/N$.

% ------------- TABLE -2 --------------- %
\begin{table}[htbp]
\vspace{-5pt}
% if use center => the blank gets larger!!!
%\begin{center}
\centering
\begin{threeparttable}[htbp]
\caption{Patient-specific Classification Performance of Backbone Networks on Different Lead Configurations\tnote{1}}
\label{table2}
% set the width of the whole chart
\setlength{\tabcolsep}{1.6mm}{
\begin{tabular}{cccccccccc}
\hline
\multirow{2}{*}{Models} & \multicolumn{4}{c}{VEB} && \multicolumn{4}{c}{SVEB} \\ 
\cline{2-5}\cline{7-10}
& Acc  & Sen & Spe & Ppr && Acc & Sen & Spe & Ppr\\
\hline
Resnet-11\tnote{2} & 98.3 & 96.1 & 98.6 & 90.1 && 97.1 & 68.0 & 98.9 & 79.2\\
Resnet-11\tnote{3} & 94.0 & 91.8 & 94.3 & 68.3 && 96.5 & 65.8 & 98.4 & 72.0\\
Resnet-11\tnote{4} & 98.6 & 94.2 & 99.2 & 93.7 && 97.2 & 69.5 & 98.9 & 79.8\\
Resnet-14\tnote{2} & 98.6 & 93.5 & \textbf{99.3} & 94.4 && 97.1 & 69.0 & 98.9 & 79.1\\
Resnet-14\tnote{3} & 97.3 & 91.5 & 98.1 & 86.6 && 97.0 & 66.3 & 98.9 & 79.7\\
Resnet-14\tnote{4} & 98.7 & 95.4 & 99.1 & 93.5 && \textbf{97.4} & 69.2 & 99.2 & \textbf{84.0}\\
Resnet-17\tnote{2} & 98.6 & 93.9 & 99.2 & 93.8 && 97.1 & 69.0 & 98.9 & 79.3\\
Resnet-17\tnote{3} & 98.0 & 92.7 & 98.7 & 90.2 && 97.1 & 66.0 & 99.0 & 80.8\\
Resnet-17\tnote{4} & 98.9 & 95.8·& 99.3 & 94.8 && 97.3 & 68.4 & 99.1 & 83.3\\
Resnet-20\tnote{2} & 98.6 & 94.9 & 99.1 & 93.2 && 97.1 & \textbf{70.4} & 98.8 & 78.2\\
Resnet-20\tnote{3} & 98.0 & 91.8 & 98.8 & 91.3 && 97.1 & 66.6 & 99.0 & 80.6\\
Resnet-20\tnote{4} & 99.0 & 95.4 & 99.5 & 96.4 && 97.3 & 67.8 & \textbf{99.2} & 83.8\\
Resnet-23\tnote{2} & 98.8 & \textbf{96.5} & 99.1 & 93.3 && 97.1 & 69.0 & 98.8 & 78.5\\
Resnet-23\tnote{3} & 98.4 & 92.3 & 99.2 & 93.9 && 97.2 & \textbf{67.5} & 99.1 & 82.4\\
Resnet-23\tnote{4} & \textbf{99.2} & \textbf{96.6} & 99.6 & 96.9 && 97.2 & \textbf{69.8} & 98.9 & 80.6\\
Resnet-26\tnote{2} & \textbf{98.9} & 95.8 & 99.3 & \textbf{95.0} && \textbf{97.2} & 68.3 & \textbf{99.0} & \textbf{80.5}\\
Resnet-26\tnote{3} & \textbf{98.5} & \textbf{93.2} & \textbf{99.3} & \textbf{94.4} && \textbf{97.2} & 65.9 & \textbf{99.1} & \textbf{82.8}\\
Resnet-26\tnote{4} & 99.1 & 95.7 & \textbf{99.6} & \textbf{96.8} && 97.3 & 69.3 & 99.0 & 81.8\\
\hline
\end{tabular}}
\begin{tablenotes}
%\item[1] The comparison results are all based on 11 common recordings for VEB detection and 14 common recordings for SVEB detection. Best results on each test scheme 	are highlighted.
\item[1] The comparison results are all based on DS2-v for VEB detection and DS2-s for SVEB detection.
\item[2] The VEB and SVEB detection results are based on the lead ML\RNum{2}.
\item[3] The VEB and SVEB detection results are based on the lead V1.
\item[4] The VEB and SVEB detection results are based on both leads.
\end{tablenotes}
\end{threeparttable}
\vspace{-10pt}
\end{table}

\section{Results}
% the sentence below should be changed...
According to the AAMI recommendations, the issues of VEB and SVEB detection are considered individually.
% tabel 3
The results for a set of control experiments are shown in Table \ref{table2} above, which relates to different ECG lead configurations and various depths of the backbone 1D Resnet.
% lead configuration
The experimental results from the models that accept both ECG leads always perform better in overall performance of VEB and SVEB detection. Most results from models using lead ML\RNum{2} surpass those from the models using lead V1 signal as input.
% depth configuration
Shallower models such as Resnet-11 and Resnet-14 generally have comparable results with those of deeper models in SVEB classification, while they perform poorly in VEB classification. Although data augmentation is implemented mainly for SVEB, VEB still has a greater information abundance than that of SVEB. The shallower models' complexity is not sufficient to converge to VEB's real distribution. Furthermore, shallower models are more likely to achieve similar results to those of deeper models in sensitivity, but do not perform equally well in precision (N is misclassified as S), indicating that the class N, with greater information richness, requires deeper networks for full encoding and decoding.
% tabel 4
Table \ref{table3} reveals the results from the models embedded with the proposed information-based channel-wise structure. The test performance indicators Sen and Ppr of SVEB are taken into consideration for model selection. The results in Table \ref{table3} are derived by the following criterion: 
% may need to recite
Set the model performance index as: $index=Sen+Ppr$, in which the sensitivity and precision are computed from the SVEB test performance.
% chosen from MCC-s 
Candidate models with top ten index values are selected from the checkpoints. The model checkpoint that provides the best MCC on SVEB is chosen for display. 
The deeper ISEnet tends to achieve better Ppr values of SVEB and VEB during the training process with the decrease of Sen as a sacrifice, which can be explained by the great disparity in terms of information quantity between the N class and S class.

% ------------- TABLE -3 --------------- %
\begin{table}[htbp]
\vspace{-5pt}
% if use center => the blank gets larger!!!
%\begin{center}
\centering
\begin{threeparttable}[htbp]
\caption{Patient-specific Classification Performance of Proposed Networks on different backbone configurations\tnote{1}}
\label{table3}
% set the width of the whole chart
\setlength{\tabcolsep}{1.6mm}{
\begin{tabular}{cccccccccc}
\hline
\multirow{2}{*}{Models} & \multicolumn{4}{c}{VEB} && \multicolumn{4}{c}{SVEB} \\ 
\cline{2-5}\cline{7-10}
& Acc  & Sen & Spe & Ppr && Acc & Sen & Spe & Ppr\\
\hline
ISEnet-11 & 99.0 & 96.2 & 99.3 & 95.1 && 97.1 & 70.2 & 98.8 & 78.9\\
ISEnet-14 & \textbf{99.4} & \textbf{97.1} & 99.7 & 97.6 && \textbf{97.7} & 70.9 & \textbf{99.4} & \textbf{87.6}\\
ISEnet-17 & 99.0 & 97.0 & 99.3 & 94.9 && 97.5 & \textbf{71.9} & 99.1 & 83.0\\
ISEnet-20 & 99.1 & 96.1 & 99.5 & 96.5 && 97.4 & 69.4 & 99.2 & 84.1\\
ISEnet-23 & 99.3 & 94.8 & \textbf{99.8} & \textbf{98.8} && 97.4 & 70.2 & 99.1 & 82.6\\
ISEnet-26 & 99.2 & 95.6 & 99.7 & 97.6 && 97.5 & 70.7 & 99.1 & 83.7\\
\hline
\end{tabular}}
\begin{tablenotes}
%\item[1] The comparison results are all based on 11 common recordings for VEB detection and 14 common recordings for SVEB detection. Best results are highlighted.
\item[1] The comparison results are all based on DS2-v for VEB detection and DS2-s for SVEB detection.
\end{tablenotes}
\end{threeparttable}
\vspace{-5pt}  % adjust the space with the context
\end{table}

% table v1
For the proposed attention model, Table \ref{table4} presents the confusion matrix of ECG beat classification test results for all the 38 records of train and test sets. The test result for 24 test records (DS2) is shown in parentheses. 
% table vi
% 245 representative beats for training dataset.
% our result cannot be compared with their results
The comparison between the chosen proposed model and the results in other works is shown in Table \ref{table5}. Our partitioning method of the benchmark training set that consists of common part and patient-specific part and test set are in consistent with the compared algorithms during the comparison implementation. Under the selected data division scheme, the base performance of the ISEnet-14, which is the best performing model in our trials, is compared with four state-of-the-art algorithms \cite{19_4}, \cite{19_7}, \cite{19_6}, and \cite{22_4} which conform to the AAMI standard. 
% need to rewrite to make sure it's the second way of data division.
 Compared with these works, the proposed model achieves higher Acc and comparable Sen in SVEB classification, and in the VEB classification issue, our model presents better results in the Sen index.
 
% model training with first 5 min data get comparable result with Kiranyaz's work in SVEB detection and derive better result in VEB.
% model training with 245 random selected NSVFQ in first dataset get better result in SVEB detection and comparable results in VEB.
% The reason backward for the difference between the two data partitioning method is that random selected heart beats covered for a longer period of time, which 

% ------------- TABLE -4 --------------- %
\begin{table}[hbp]
\vspace{-5pt}
% if use center => the blank gets larger!!!
%\begin{center}
\centering
\caption{Confusion matrix of ISEnet-14 testing on MIT-BIH records}
\label{table4}
% set the width of the whole chart
\setlength{\arraycolsep}{4pt}
% using -size keyword to control the fontsize in table
\scriptsize
\setlength{\tabcolsep}{0mm}{
\begin{tabular}{ccccccc}
\hline
\multirow{2}{*}{Label} & \multicolumn{5}{c}{\quad \quad \quad \quad Prediction}\\ 
\cline{2-7}
\specialrule{0em}{0.5pt}{0.5pt}
& N & S & V & F & \ Q \  & ${\sum}$\\
% insert transparent line to enlarge the space
\specialrule{0em}{0.5pt}{0.5pt}
\hline
\specialrule{0em}{0.5pt}{0.5pt}
N &77950 (49526) &507 (306) &377 (174) &170 (162) &\ 0 (0) \ &79004 (50168)\\
S &934 (885) &1904 (1840) &77 (76) &11 (11) &\ 0 (0) \ &2926 (2812)\\
V &284 (193) &32 (19) &6543 (5463) [6572]&52 (49) [23]&\ 0 (0) \ &6911 (5724)\\
F &46 (44) &1 (1) &48 (48) [19] &698 (696) [727]&\ 0 (0) \ &793 (789)\\
Q &4 (4) &0 (0) &9 (3) &0 (0) &\ 0 (0) \ &13 (7)\\
$\textstyle\sum$ &79218 (50652)&2444 (2166) &7054 (5764) &931 (918) &0 (0) &89647 (59500)\\
\specialrule{0em}{0.5pt}{0.5pt}
\hline
\end{tabular}}
\end{table}

\vspace{-10pt}
\section{Discussion}
\subsubsection{More channels can be better}
The results listed above show that utilizing dual-lead ECG as the input benefits the ECG classification tasks. This is also confirmed in literature \cite{m_7} and \cite{m_8}.
% need to check 
Fig. \ref{figure6} is used to characterize the model complexity and the overall model performance in MCC index on the ECG classification task in the ablation study.
Models that accept dual leads as input surpass the performance of those accept single lead in ECG classification tasks with only a small increase in model complexity. 
% using \vspace & \hspace to control the interval
\begin{figure*}[htbp]
\centering
% 1
\subfigure[ISEnet-14 Block\_1 Unit\_1.]
{\label{e1}\includegraphics
[width=1.8in, height=1.17in]{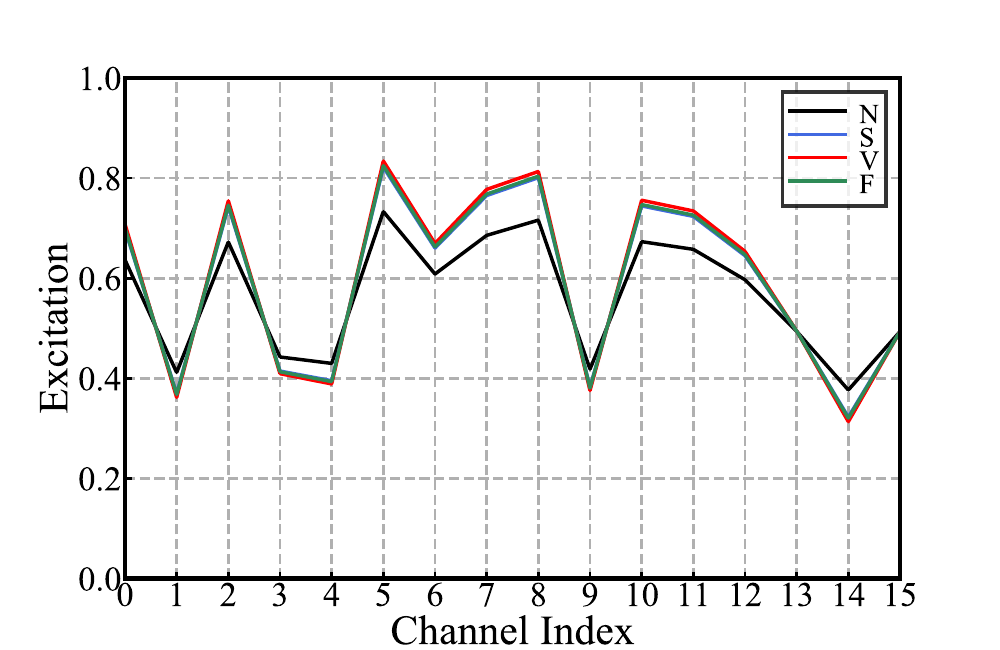}}
%\vspace{0in}
\hspace{-0.12in}
% 2
\subfigure[ISEnet-14 Block\_2 Unit\_1.]
{\label{e2}\includegraphics
[width=1.8in, height=1.17in]{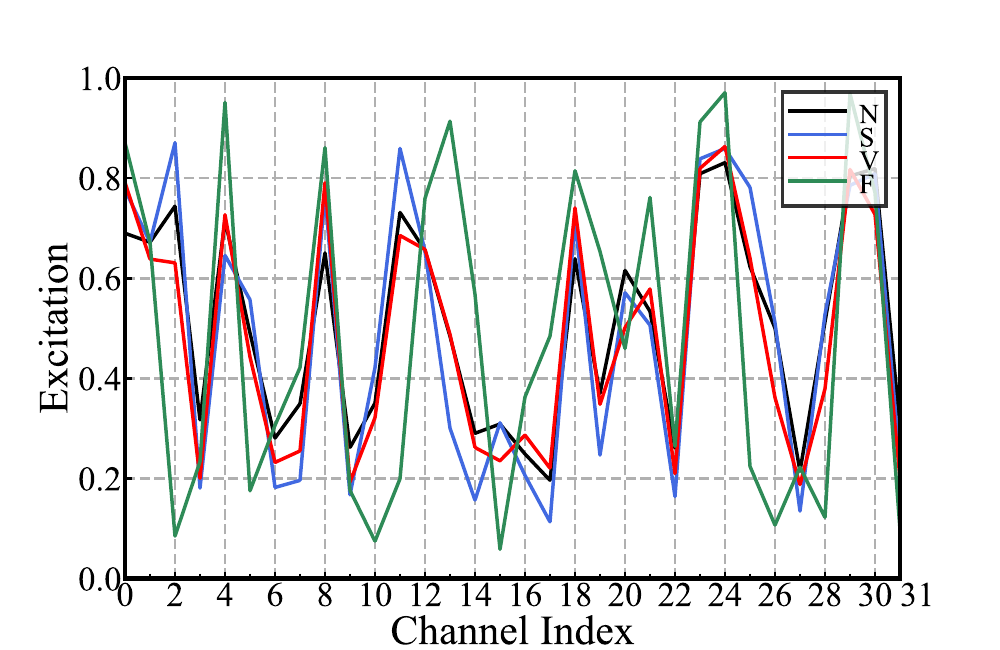}}
%\vspace{0in}
\hspace{-0.15in}
% 3
\subfigure[ISEnet-14 Block\_3 Unit\_1.]
{\label{e3}\;\includegraphics
[width=1.8in, height=1.17in]{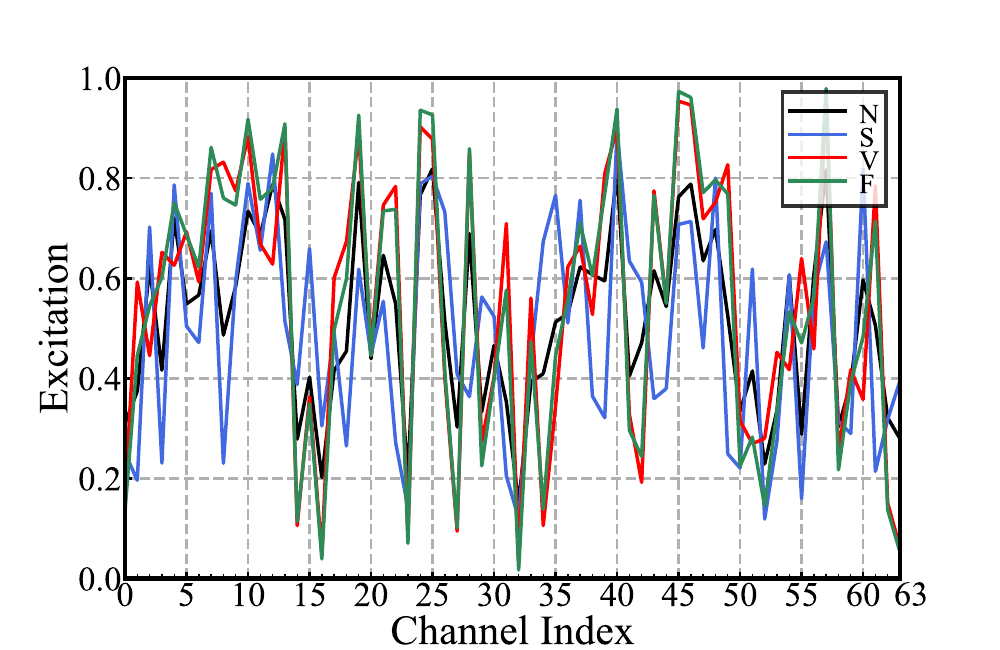}}
%\vspace{0in}
\hspace{-0.13in}
% 4
\subfigure[ISEnet-14 Block\_4 Unit\_1.]
{\label{e4}\includegraphics
[width=1.8in, height=1.17in]{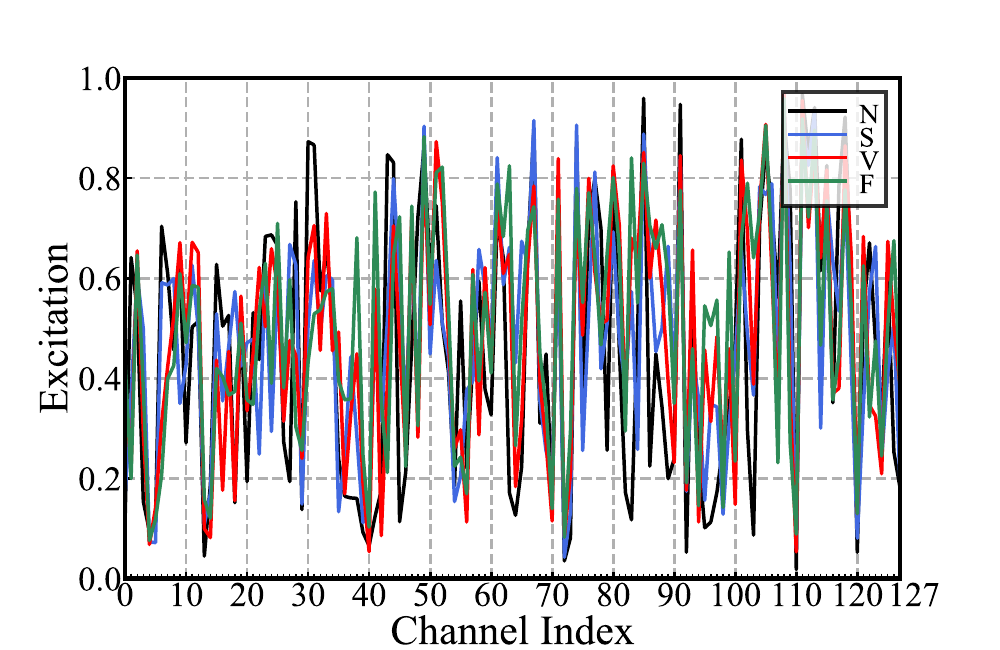}}
%\vspace{0in}
\caption{Excitation value induced by structure P in Figure \ref{figure3} of the various blocks of ISEnet-14 on N, S, V, F samples from the test set of mitdb. See Table \ref{table1} for details of ISEnet-14 block and unit configurations.}
% use \subref{} to ref the sub-pic
\label{figure7}
\vspace{-10pt}  % adjust the space with the context
\end{figure*}
Reasonably increasing the dimension of the input features on the premise of intensively debugging the hyper-parameters and appropriately applying training techniques will greatly improve the performance of the model and resist the tendency of over-fitting.

% --------- table -5 -------------%
\begin{table}[htbp]
\vspace{-0.cm}
\centering
% three part table is for usage of \tnote{xxx}
\begin{threeparttable}[htbp]
\caption{VEB and SVEB Classification Performance and Comparison With Four State-of-the-art literatures}
\label{table5}
% set the width of the whole chart
%\setlength{\arraycolsep}{5pt}
\setlength{\tabcolsep}{1.4mm}{
\begin{tabular}{cccccccccc}
\hline
\multirow{2}{*}{Models} & \multicolumn{4}{c}{VEB} && \multicolumn{4}{c}{SVEB}\\ 
\cline{2-5}\cline{7-10}
& Acc  & Sen & Spe & Ppr && Acc & Sen & Spe & Ppr\\
\hline
Hu et al. & 94.8 & 78.9 & 96.8 & 75.8 && N/A & N/A & N/A & N/A \\
Jiang and Kong\tnote{2} & 98.8 & 94.3 & \textbf{99.4} & 95.8 && \textbf{97.5} & 74.9 & 98.8 & 78.8 \\
Jiang and Kong\tnote{3} & 98.1 & 86.6 & 99.3 & \textbf{93.3} && 96.6 & 50.6 & \textbf{98.8} & \textbf{67.9} \\
Ince et al.\tnote{2} & 97.9 & 90.3 & 98.8 & \textbf{99.2} && 96.1 & \textbf{81.8} & 98.5 & 63.4 \\
Ince et al.\tnote{3} & 97.6 & 83.4 & 98.1 & 87.4 && 96.1 & 62.1 & 98.5 & 56.7 \\
Ince et al.\tnote{4} & 98.3 & 84.6 & 98.7 & 87.4 && 97.4 & \textbf{63.5} & 99.0 & 53.7 \\
Kiranyaz et al.\tnote{2} & 98.9 & 95.9 & \textbf{99.4} & 96.2 && 96.4 & 68.8 & \textbf{99.5} & \textbf{79.2} \\
Kiranyaz et al.\tnote{3} & \textbf{98.6} & 95 & 98.1 & 89.5 && 96.4 & \textbf{64.6} & 98.6 & 62.1 \\
Kiranyaz et al.\tnote{4} & \textbf{99.0} & 93.9 & \textbf{98.9} & \textbf{90.6} && \textbf{97.6} & 60.3 & \textbf{99.2} & \textbf{63.5} \\
ISEnet-14\tnote{2}& \textbf{98.9} & \textbf{96.7} & 99.2 & 94.5 && 97.1 & 68.0 & 98.9 & 79.1\\
ISEnet-14\tnote{3}& 98.4 & \textbf{95.2} & \textbf{98.7} & 89.1 && \textbf{96.8} & 63.3 & 98.4 & 66.9 \\
ISEnet-14\tnote{4}& 98.6 & \textbf{94.5} & \textbf{98.9} & 88.2 && \textbf{97.6} & 63.1 & 98.7 & 62.8\\
\hline
\end{tabular}}
\begin{tablenotes}
\item[2] The comparison results are based on DS2-v for VEB detection and DS2-s for SVEB detection.
\item[3] The VEB and SVEB detection results are compared for DS2 only.
\item[4] The VEB and SVEB detection results of the proposed models for testing part of DS1 and DS2.
\end{tablenotes}
\end{threeparttable}
% attention that: using \vspace{} to force the space between the table and the context
\vspace{-5pt}
\end{table}

\subsubsection{Comprehension of the layers' function through excitation values} 
To study the functions of different ISE-blocks embedded in the middle of the backbone networks, the ISEnet-14 is selected as a base and tested using randomly selected 500 samples from each class except Q in the test datasets. Then, we plot the average excitation values characterized by matrix P in the Fig. \ref{figure3} of all the samples, as shown in Fig.\ref{figure7}. The above sample selection process is repeated multiple times to ensure that the pattern of the excitation values we obtain do not depend on the contingency of selecting the test data.
% need to change!
Two observations are made from the visualization of the proposed structures. 
As shown in subfigure a in Fig.\ref{figure7}, the excitation values of Block1 Unit1 distributes between 0.313 and 0.834, with merely six of the sixteen channels lying under 0.5, revealing that most channels values characterizing the general features are preserved in shallower layers of the network. In the deeper depths of the model, the range of the excitation values stretches to 0.0171 -- 0.979, which indicates significant suppression and excitation of different channels on the feature map. That is, the main function of the shallow layers is to extract general information with few differences among categories, and the deeper layers are majorly responsible for the compression and integration of the relevant information towards the sample types.
Furthermore, for each embedded ISE-block (a -- b) in Fig.\ref{figure7}, channel-wise activation values are adopted to calculate the mean square deviation defined below for any two classes:
% mcc
\begin{equation}
\begin{split}
MSD_{m, n} = \frac{M}{C}\sum_{i=0}^{C-1}{(m_i-n_i)^2}
\end{split}
\label{e14}
\end{equation}
where C denotes the number of channels depicted in Fig.\ref{figure7}, M denotes the multiplier factor and is set to 100, and m and n stands for the two classes involved in the computation.
For N-S, the MSD values from subfigure a to subfigure d are 0.329, 0.821, 2.425, and 3.764.
For N-V: they are 0.442, 0.545, 2.909, and 4.052. 
For N-F, S-F and V-F, the MSD values computed from the block2 are the highest.
For S-V, the MSD values from the block3 are the highest.
\iffalse The proposed models consists of two parts: the Resnet backbone and the feature map importance recalibration embedments.\fi Ignoring the effect of the back-propagation algorithm on model convergence, we argue that the MSD values derived above partly characterize the relationship between the information flow on the selected two classes and the ability for model segments to encode and decode, which sliced from the input layer to the layer that used to calculated MSD values.
% experimental results of the using msd as index
Taking $MSD_{v,f}$ as an case, we truncate the trained ISEnet-14 model at the end of block2 and append it with a convolutional layer and a global average pooling layer to form an model towards $V$ and $F$ classification. The parameters in the segmented part are fine-tuned with a smaller learning rate, and the appended layers are trained with the selected two classes. The refined results derived from the two combined models are shown in Table \ref{table4} enclosed in square brackets.
% need experiments results to confirm this ...
The above phenomenon and its analysis results reveal that for the ECG classification task, the smallest network segment that is advantageous for the distinction between the two types of ECG classes can be obtained by calculating the MSD values. The reasonable use of the network segment selected according to the MSD index can improve the accuracy of the ECG classification issue and the generalization ability of the corresponding model.

% ---- mcc performance chart ---- %
\begin{figure}[htb]
\label{figure6}
\centering
\subfigure[SVEB detection results in Matthews correlation coefficient of different models with related FLOPs.]{\label{v0}
\includegraphics[width=3.5in]{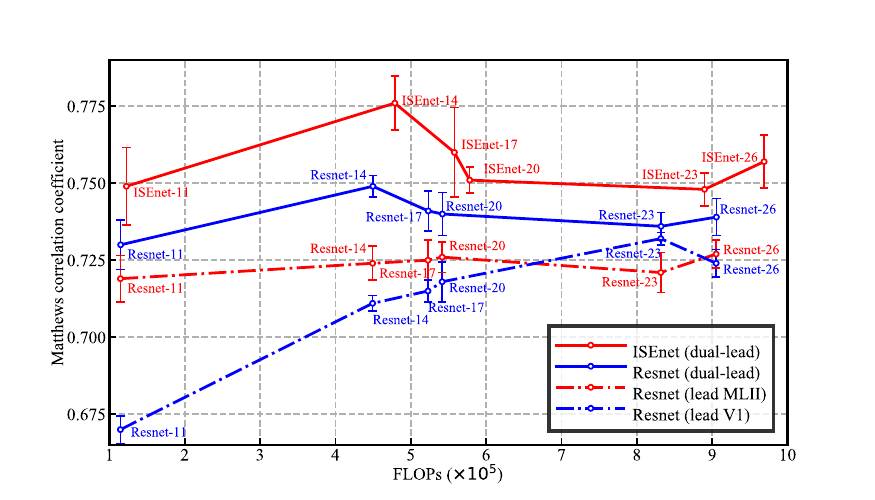}} 
\\
\subfigure[VEB detection results in Matthews correlation coefficient of different models with related FLOPs. The y scale is processed using the logit function for better display of the contrast of different models' performances.]{\label{v1}
\,\includegraphics[width=3.49in]{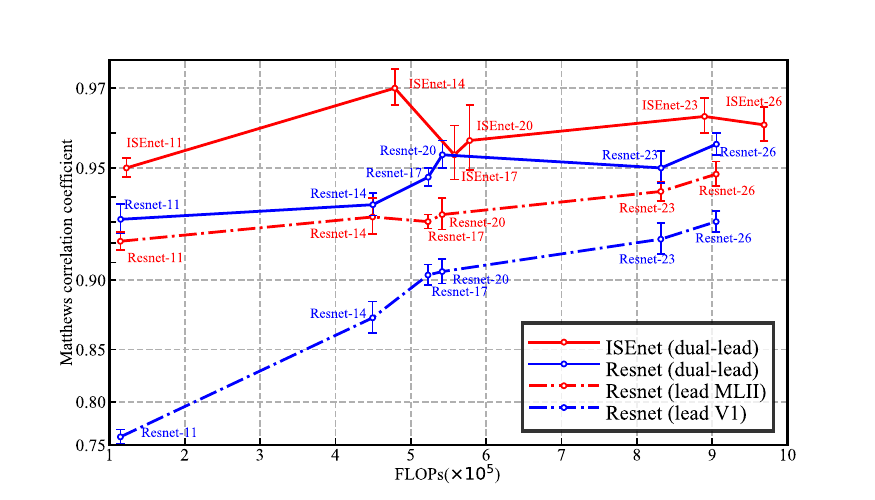}} 
\caption{The relationship between model complexity and model performance using Matthews correlation coefficient as an index in our ablation study. The ranges of the error bars are calculated by $M \pm std$, where $M$ and $std$ are mean value and standard deviation from the results of 6 repetitive experiments of the same model.}
\label{figure6}
% use \subref{} to ref the sub-pic
\vspace{-5pt}  % adjust the space with context
\end{figure}

% may need to changed...
\subsubsection{Backbone structures mining}
\begin{figure}[htbp]
\vspace{-10pt}
% 1
\hspace{-0.8pt}  % adjust the centering props.
\subfigure[\scriptsize{Resnet}]
{\label{v1}\includegraphics
[height=1.0in]{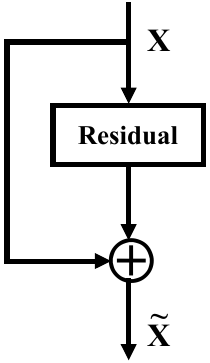}}
%\vspace{0in}
\hspace{0.05in}
% 2
\subfigure[\scriptsize{Standard-block}]
{\label{v2}\quad \includegraphics
[height=1.0in]{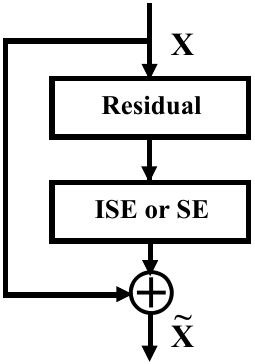}}
%\vspace{0in}
\hspace{0.1in}
% 3
\subfigure[\scriptsize{Pre-block}]
{\label{v3}\;\includegraphics
[height=1.0in]{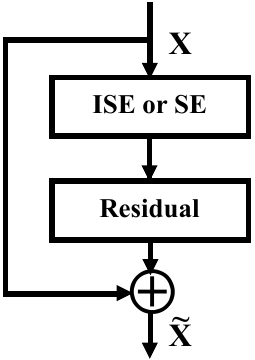}}
%\vspace{0in}
\hspace{0.1in}
% 4
\subfigure[\scriptsize{Identity-block}]
{\label{v4}\includegraphics
[height=1.0in]{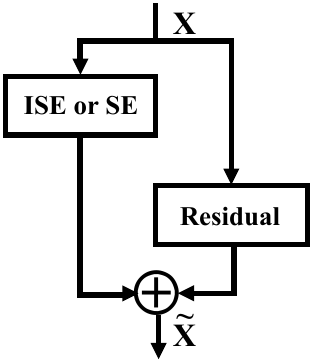}}
\vspace{-10pt}
\caption{Resnet block together with standard ISE block and its variants.}
% use \subref{} to ref the sub-pic
\label{figure8}
\vspace{-10pt}
\end{figure}

The classic resnet-v2 structure and three variants of the ISE-block are shown in Fig. \ref{figure8}. In this study, the standard structure of the ISE-block is chosen. We made preliminary attempts on the other two variants, and they could also perform well in the ECG classification task. This reveals that the ISE-block can be embedded as a general unit in the network. \iffalse According to the result presented above, models embedded with ISE-blocks always achieve superior results in terms of the four assessment indicators.\fi Other network backbones and ISE-variants may be able to achieve similar results in solving the classification task, which is beyond the scope of this study.

% multi fig

\section{Conclusion}
% A conclusion might elaborate on the major findings and significance of the work or suggest applications and extensions. Do not exceed 300 words for the conclusion section.
% ----- advantages ----- %
The backbone network structures adopted are a ``one-stage'' training strategy, which can realize automatic feature extraction to achieve better robustness and higher adaptability in ECG classification issues compared with methods involving manually extracting features. 
\iffalse
% need to recite >>>>>>>>>>>>>>>>>>
Our algorithm has a little dependence on data preprocessing. 
% are u sure ?????????????????????
Compared with most algorithms, our designed network architecture is less dependent on quantity of samples in training set.
\fi
A set of data partitioning and reconstruction schemes is proposed for applying dual-lead signals in the mitdb for the neural networks. The proposed scheme conforms to the AAMI standard and fully satisfies the requirements of deep neural network algorithms.
% need to rethinking or rewrite
With reference to the information bottleneck theory, we derive the possibility that applying statistics in the feature maps of the backbone model helps quantify the ``valid information quantity'', which is used as an index to monitor the neural network training quality and to help comprehend the relationship between the model structures and its functionalities.
The perspective that utilizing multi-channel signals can help train better generalized deep networks with superior performance in ECG classification tasks is proven through a series of control experiments. 
We design a structure based on a channel attention mechanism, which can effectively realize the channel-wise weights recalibration (excitation and suppression). The proposed model has the potential to use more leads for the analytical diagnosis of more complex heart diseases.

% ----- drawbacks ----- %
However, it should be noted that the ECG signal is a time sequence, while the proposed network structure fails to utilize the characteristics of long-term time series of the signal, which may lead to the failure of the analysis of heart diseases characterized by episode-level features. In addition, the performance of our model is limited by the quality of the beat-by-beat segmentation algorithm and the signal preprocessing schema. Due to the neural network's black-box nature and the simplifications made in the design of the ISE structures, it is not yet possible to strictly prove the validity of the designed structures in theory.

% ----- todo ----- %
In the future, we will use RNN and its variants as the backbone to model the long-term ECG signals, and to figure out the dependency between the more complicated heart diseases with its lead configurations. We will go further to prove the argument that integrating the MSD index with the network model design will benefit the performance of the classification model.

\section*{Appendices}
% ------- Go deeper in ISE-block ------ %
\subsubsection{Go deeper in the ISE-block}
\setlength{\parskip}{0em} % 
\setlength\abovedisplayskip{6pt}  % up-skip
\setlength\belowdisplayskip{6pt}  % down-skip
% -- the information bottleneck theory --%
The idea behind the proposed ISE-block comes from the expression of the hidden layers in neural networks based on the information bottleneck theory in \cite{q_1} and \cite{q_2}. We argue that neural networks can be interpreted as information encoding and decoding, which are structures that integrates and process information. \iffalse They can be divided into the fusion structure of encode and decode, corresponding to the process of information compression and prediction.\fi These two functions are implemented simultaneously in any hidden layer during the training of the neural network. Schwartz-Ziv and Zaslavsky provides a good mathematical explanation of the above viewpoint \cite{q_1}: Let $T_i$ represents the $i$-th hidden layer, and the $T_i$ is uniquely mapped to a point in the information-plane with coordinates $(I(X;T_i), I(T_i;Y))$. In \cite{q_2}, Tishby et al. argue that the efficiency of the network's hidden layers can be quantified by the amount of information it retains on the input variable. Thus, we consider trying to quantify the flow of information in a neural network by grafting a structure in a neural network. The quantified results are applied to different channels of the original feature maps to enhance the fusion of neural network channel features, to achieve better results in multi-channel ECG classification modeling tasks.

% ----- i(x, y) definition ----- %
We use the mutual information to quantify the valid information of $T_i$, which can be thought of as the amount of information about another random variable contained in a random variable. 
The mutual information of any two random variables X and Y can be defined as:
% ----- mutual information equation ----- %
\begin{equation}
\begin{small}  % smaller the characters of equation
%\begin{split}  % automatically split the equation
I(X;Y)=D_{KL}[p(x, y)\|p(x)p(y)]=H(X)-H(X|Y) \label{6}
\end{small}  % should not using \small front of equation!
\end{equation}
where $p(x, y)$ is the joint distribution of $x$ and $y$, $D_{KL}(p\|q)$ is the KL divergence, and $H(X)$ and $H(X|Y)$ are the entropy and conditional entropy.
% ---- derive information embeddings for DNN ---- %
First, we derive the expression of the ``valid information expression'' based on the structure of  deep neural network (DNN). 
Let $X$ be the input of the DNN, and let $Y$ be the random variable identified by the label distribution, and $T_i$ represents the whole $i$-th hidden layer, which can be treated as a single random variable and represented by $P(T_i|X)$ and $P(Y|T_i)$ distributions \cite{q_1}. Thus, $I(T_i;Y)$ is the quantity of information about the random variable $Y$ contained in $T_i$.

% ---- derive information embeddings for CNN ---- %
The ``valid information expression'' of a CNN is derived in the following part. The convolution operation is a reconstructed sparse form of the fully-connected structure in a DNN. For a given feature map $T_i$ with $3\times 3$ nodes named $t_{i0}, t_{i1}\dots t_{i8}$, a kernel $K_i$ with $2\times 2$ nodes named $k_{i0}, k_{i1}, k_{i2}$ and $k_{i3}$, and the resulting output with 4 nodes named $o_{i0}, o_{i1}, o_{i2}$ and $o_{i3}$, taking the output node $o_{i0}$ as an instance, the convolution ($stride=1$) on the input $T_i$ can be transformed into the following equation:
% ----- conv into dnn ----- %
\begin{equation}
\begin{split}
\bm{o}_{i0} = t_{i0}\times k_{i0}+t_{i1}\times k_{i1}+t_{i2}\times 0+t_{i3}\times 0+t_{i4}\\\times k_{i2}+t_{i5}\times k_{i3}+t_{i6}\times0+t_{i7}\times0+t_{i8}\times0  \label{e7}
\end{split}
\end{equation}
For the $i$-th hidden layer in a CNN model expressed as $T_{i}$, let $j$ denote one of its total $c$ channels; thus, $T_{i}[j]$ is the $j$-th feature map in $T_{i}$, and it is derived by the convolution of the $j$-th kernel and the $T_{i-1}[j]$. The output feature map of a certain kernel is combined by weight sharing and sparse connection, and is directly computed through element-wise addition.
% which may lead to redundant or erroneous information fusion and can be a reason for the trickiness of network training convergence.
Given the relationship between CNN and DNN derived above, we argue that each $T_{i}[j]$ is a single random variable, and the ``valid information expression'' can be derived as $I(T_{i}[j],Y)$.
% ----- theoretical problem is hard ----- %
Thus, we can use the expression to get the relative ``importance score'' between different channels of the same layer of the convolutional neural network. However, in the training process of CNN models, the distribution of a hidden layer $P(T_i[j])$ is hard to calculate, and it is not possible to flush all the training data into the model to approach the true distribution, which motivates us to seek an alternative that is both computationally friendly and close to the original mathematical principle.
% --- our simplifications to represent I(x, y) ---%
We made several simplifications for the expression: $I(T_{i}[j];Y)=H(T_{i}[j])-H(T_i[j]|Y)$, in which $H(T_i[j])$ and $H(T_i[j]|Y)$ are both hard to estimate, we adopt its symmetrical form: $I(Y;T_{i}[j])=H(Y)-H(Y|T_{i}[j])$, in which the $H(Y)$ is a constant and is fully determined by the label distribution. Thus the expression can be simplified by cutting off its constant part. Given $H(H|X)=\sum_xp(x)H(Y|X=x)$, the information expression can be computed as:
% ----- H(y|x) ----- %
\begin{equation}
\begin{small}
H(Y|T_i[j])=-\sum_{t_i[j], y}p(t_i[j], y)\ log\,p(y|t_i[j])\label{e8}
\end{small}
\end{equation}
where $t_i[j]$ denotes the observation value of the random variable $T_i[j]$.
A single forward propagation of a neural network can only yield one observation $\hat{t_i[j]}$ in the $T_i[j]$ distribution. The distribution of $Y$ can be approximated by observing the label distribution. However, we still cannot directly solve the two probability values in (\ref{e8}). Through the back propagation in training each neural network, the distribution of a certain hidden layer converges to its true distribution, upon which we can implement quantification and optimization operations on. Moreover, the expression of (\ref{e8}) provides us an approximate way of measuring the quantity of the information flow. That is, we can provide channel recalibration information for the training process by observing the internal numerical statistics of $\hat{t_i[j]}$ within each inference. Finally, we use two FC layers as calibration structures for channel-wise learning to compensate for the simplifications above.

%\section*{Acknowledgment}
%The author would like to thank for the reviewer for advices.
%Sponsor and financial support acknowledgments are placed in the unnumbered footnote on the first page, not here.

% ------ references section ------ %
\vspace{-10pt}
\bibliographystyle{IEEEtran}
\bibliography{reference.bib}

\end{document}